\documentclass[aip,jcp]{revtex4-1}
\draft 
\usepackage{hyperref}
\usepackage{graphicx}
\usepackage{tabularx} 

\begin{document}
\title{\Large Evolution of ion track morphology in \textit{a}-SiN$_\mathrm{\textit{x}}$:H by dynamic electronic energy loss} 
 
\author{R K Bommali}
\email{ravibommali06@gmail.com}
\affiliation{Department of Physics, St. Xaviers College Ranchi}
\author{S Ghosh}
\email{santanu1@physics.iitd.ac.in}
\author{Harsh Gupta}

\affiliation{Department of Physics,Indian Institute of Technology Delhi}
\author{Rafael Perez Casero}
\affiliation{Department of Applied Physics, Universidad Autonoma de Madrid}
\author{P Srivastava}
\affiliation{Department of Physics,Indian Institute of Technology Delhi}
\date{\today}

\begin{abstract}
Amorphous hydrogenated silicon nitride (\textit{a}-SiN$_\mathrm{\textit{x}}$:H) thin films irradiated with 100 MeV Ni$^{7+}$ results in the formation of continuous ion track structures  at the lower fluence of $5\times{10^{12}}$ ions/cm$^2$ whereas at higher fluence of $1\times{10^{14}}$ ions/cm$^2$ the track structures fragment into discontinuous ion track like structures . The observation of the discontinuous ion track like structures at the high fluence of $1\times{10^{14}}$ ions/cm$^2$ shows clearly that higher fluence irradiation may not always lead to dissolution of the microstructure formed at lower fluence. The results are understood on the basis of a dynamic electronic energy loss (S$_{e}$) in the course of irradiation resulting from the out-diffusion of hydrogen from the films and a continuous increase in density of \textit{a}-SiN$_\mathrm{\textit{x}}$:H films
\end{abstract}


\maketitle 

\section{Introduction}
Amorphous hydrogenated silicon nitride (\textit{a}-SiN$_\mathrm{\textit{x}}$:H) deposited by PECVD is a precursor material to obtain Si quantum dots (QDs) embedded in a silicon nitride matrix. Conventionally, Si rich films are subjected to various post deposition treatments viz. furnace annealing, flash lamp annealing, irradiation with swift heavy ions etc., to effect the phase separation of the Si and the silicon nitride phase. Among the above post treatments swift heavy ion (SHI) irradiation of \textit{a}-SiN$_\mathrm{\textit{x}}$:H thin films is of potential interest to achieve the much desired control over the size and spatial distribution of the embedded quantum dots. The localized and transient energy deposition mechanism inherent in SHI irradiation  \cite{toulemonde2000}$ ^{,} $ \cite{toulemonde2011}$ ^{,} $\cite{agullo2016}, within a small cylindrical volume around the ion path raises the possibility of controlling the size and spatial distribution of QDs. The growth of QDs along the ion track walls has been demonstrated in the past by Antonova et al. for Si rich silicon oxide system \cite{antonova2009}. Further, in our previous work\cite{bommali2015} we have reported formation of discontinuous ion tracks in \textit{a}-SiN$_\mathrm{\textit{x}}$:H, wherein by an excitation dependent photoluminescence study the formation of amorphous Si QDs is reported. The aforementioned control of size and spatial distribution of QDs is not possible with conventional post-treatments. Another reported approach of controlling the size and spatial distribution of QDs by growing superlattices of Si/ Si$_3$N$_4$ to achieve size control was found to lead to the formation of multiple parallel interfaces \cite{zelenina2013} that make  the extraction of the QD luminescence from the films difficult. 
	
	SHI irradiation promises a reproducible and effective approach to achieve the aforementioned objectives. For example the size of the QDs can be potentially controlled by ion velocity \cite{meftah1994} and type, which determines the cylindrical volume of affected region around the ion trajectory \cite{ridgway2015}, known as the ion track. The irradiation induced defects on track walls may provide nucleation sites \cite{antonova2009} for the growth of QDs, so that the irradiation fluence may determine the separation between adjacent QDs. However, the technique can be beneficially used for the particular case of \textit{a}-SiN$_\mathrm{\textit{x}}$:H, only if the interaction of SHIs with it and the concomitant processes are fully understood. Recently, the interest in silicon nitride based materials \cite{murzalinov2018}$ ^{,} $\cite{van2020}$ ^{,} $\cite{kitayama2015} has surged and they are being investigated for track formation, as silicon nitride offers a higher density contrast \cite{mota2018} in the over-dense and under-dense regions of the ion tracks as compared to SiO$_2$. 
	Coming to the case of hydrogenated silicon nitride (\textit{a}-SiN$_\mathrm{\textit{x}}$:H), it must be understood that \textit{a}-SiN$_\mathrm{\textit{x}}$:H is quite different from the materials being dealt with in the vast majority of the reports \cite{ridgway2013}$ ^{,} $\cite{kluth2014} studying ion track formation/ion matter interaction in the SHI regime. The hydrogen incorporated in \textit{a}-SiN$_\mathrm{\textit{x}}$:H, imparts it a structural metastability, namely, that these films undergo large compaction \cite{bommali2018} (10-20 \%) in their thickness in the course of irradiation. The compaction results from an out-diffusion of molecular hydrogen. As a result of which the film density and microstructure are continually evolving with irradiation fluence. In other words, the fundamental parameter, electronic energy loss (S$_e$)  which determines the nature of ion matter interactions changes continuously in the course of irradiation. This kind of continuous evolution deviates from the standard understanding of the sequence of ion matter interaction, which starts with damage creation resulting in ion track formation followed by overlap of the cylindrical damaged zones or “ion tracks” as the fluence exceeds a certain critical fluence. In contrast, in the present case the formation of ion tracks in the track overlap regime (at fluence of $1\times{10^{14}}$ ions/cm$^2$) is evidenced. In the present work we therefore attempt to understand the processes that determine the ion track evolution in \textit{a}-SiN$_\mathrm{\textit{x}}$:H films.

\section{Experimental Details}
\textit{a}-SiN$_\mathrm{\textit{x}}$:H thin film was deposited on, p-type Si (100), using PECVD (SAMCO PD-2S) technique which uses rf (13.56 MHz) power to maintain a glow discharge and produce ionised species necessary for thin film growth. Silane (SiH$_4$) (4\% in Ar) and ammonia (NH$ _{3} $) were used as precursor gases. During deposition the substrate temperature, chamber pressure and plasma power were maintained at 200 $ ^{\circ} $C, 1.7 mbar and 25 W, respectively. \textit{a}-SiN$_\mathrm{\textit{x}}$:H thin films having dimension 10$ \times $10 mm$ ^{2} $ were irradiated in a direction normal to the film surface with 100 MeV Ni$ ^{7+} $ ions using a 15 UD Pelletron accelerator at the Inter University Accelerator Centre, New Delhi, India. The films were irradiated at room temperature under a chamber pressure less than 10$ ^{-7} $ mbar. The beam was magnetically scanned in order to irradiate the film surface uniformly with fluences of $5\times{10^{12}}$ ions/cm$^2$ and $1\times{10^{14}}$ ions/cm$^2$. The film compositions were determined using Rutherford backscattering spectrometry (RBS). Cross- Sectional Transmission Electron Microscopy (XTEM) measurements were carried out using a Philips CM-200 microscope, operating at 200 kV, to examine formation of microstructures in the as-deposited and irradiated films. XTEM samples were prepared using the standard procedure involving mechanical polishing and ion milling. The experimental details of photoluminescence (PL) and elastic recoil detection analysis (ERDA) are similar to those provided elsewhere \cite{bommali2014}. RBS was carried out using 2.3 MeV He$^{4+}$ ions at a backscattering angle of 170$ ^{\circ} $. The detector solid angle was 2.5 milli-steradians and its resolution was 23 keV.
\begin{table*}
\caption{\label{Tab1}The composition, density and hydrogen concentration with irradiation fluence of \textit{a}-SiN$_\mathrm{\textit{x}}$:H thin films. The percentage compaction ($\Delta$t\%) of the films at the fluence of $1\times{10^{14}}$ ions/cm$^2$ relative to as-deposited thickness (t) is also depicted. The sample labeled B is the sample considered in the present work, while R and G are films reported previously\cite{bommali2015}}
\begin{tabularx}{\textwidth}{ p{1.6cm}  p{1.6cm}  p{2.4cm}  p{2cm}  p{2.5cm}  p{2.5cm} p{0.4cm} p{1cm} p{1cm}}
\hline
		{}&{}&{as-deposited  }&\multicolumn{3}{c}{H-Concentration (atoms/nm$  ^{3}$)}&{}&{}&{}\\
		\cline{4-6}
		{sample}&(N/Si)&(Density\newline gm/cc)&as-deposited&$5\times{10^{12}}$ \newline (ions/cm$^2$&$1\times{10^{14}}$) \newline (ions/cm$^2$)&{}&\ \ t&$\Delta$ t \%\\
		\hline	\hline	
		B&1.4&2.2&44&35&14&{}&253&15\\
		G&1.0&2.1&33&...&...&{}&115&9.4\\
		R&0.9&1.9&24&...&...&{}&109&17\\ 
		\hline	
		\end{tabularx}
		\vspace{6pt}
\end{table*}

\section{Results and discussion}
The stoichiometry ratio (N/Si), H content and the compaction of \textit{a}-SiN$_\mathrm{\textit{x}}$:H  thin films before and after irradiation at $5\times{10^{12}}$ and $1\times{10^{14}}$ ions/cm$^2$ are listed in Table \ref{Tab1}. The thin film labelled B is in consideration in the present paper. Compositions of films R and G, from our past publication are also depicted for comparison of the initial hydrogen content and resulting percentage compaction ($\Delta$ t \%) after irradiation. It must be noted that the compactions have been only determined after the irradiation at $1\times{10^{14}}$ as the compaction after the low fluence irradiation of  $5\times{10^{12}}$ ions/cm$^2$ is insignificant \cite{bommali2017}.
\begin{figure}[h]
\centering
\includegraphics[scale=0.6]{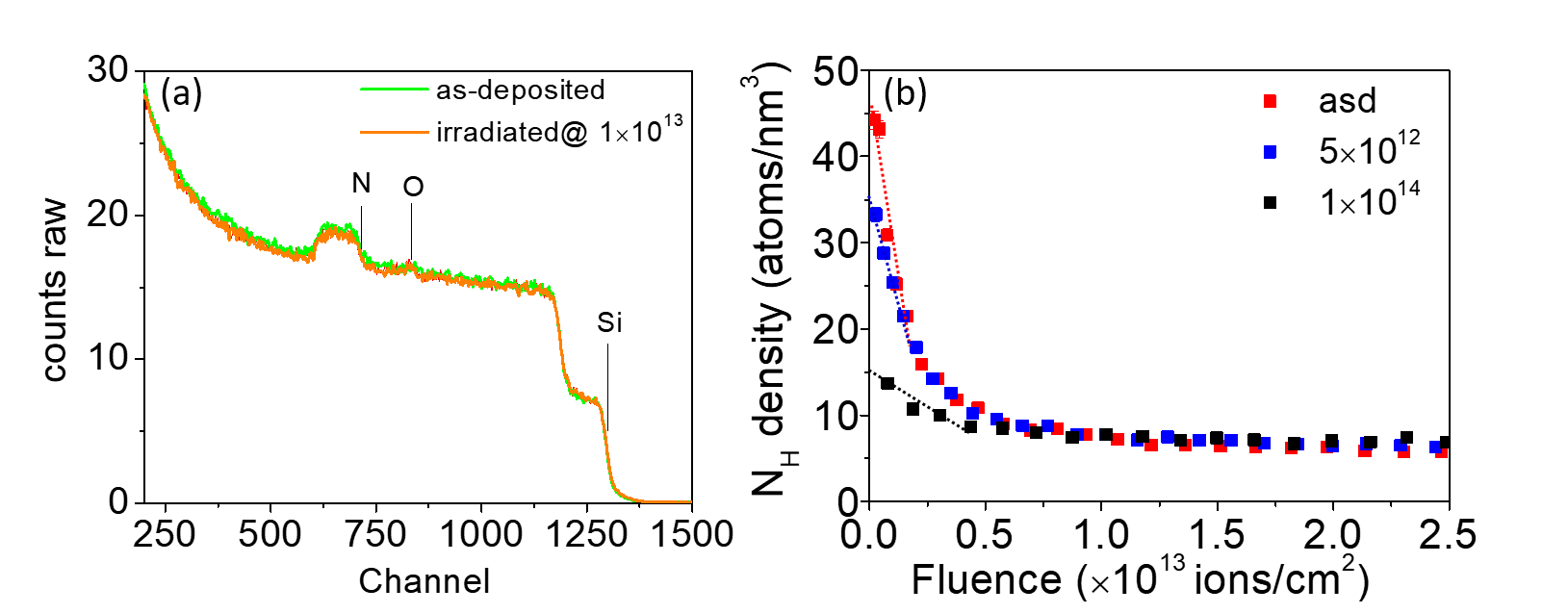}
\caption{\label{Fig1}(a) RBS spectra for the as-deposited and irradiated film showing negligible composition changes in the films during irradiation (b) ERDA data showing the hydrogen content in the film for as-deposited and films irradiated at $5\times{10^{12}}$ and $1\times{10^{14}}$ ions/cm$^2$ .}
\end{figure}

The composition of the as-deposited film has been determined by RBS. The N/Si ratio of 1.44, indicates a nitrogen rich film under consideration. Further, if we normalise\cite{bommali2018} the areal densities obtained from RBS  and ERDA according to the thickness of the films, we obtain the atomic concentration of Si, N and H as 27, 39 and 44 atoms/nm$^3$ respectively. The high concentration of hydrogen for the present nitrogen rich film is consistent with our earlier findings \cite{bommali2014}. Further, assuming that the N and Si content of the films remain constant in the course of irradiation, so that the densification of the films can be ascribed to the compaction due to hydrogen out-diffusion only. This insignificant change in composition of the films is corroborated by the RBS spectra depicted in figure \ref{Fig1}a for the as-deposited and irradiated film at $1\times{10^{13}}$ ions/cm$^2$.

\begin{figure*}[t]
\centering
\includegraphics[scale=0.25]{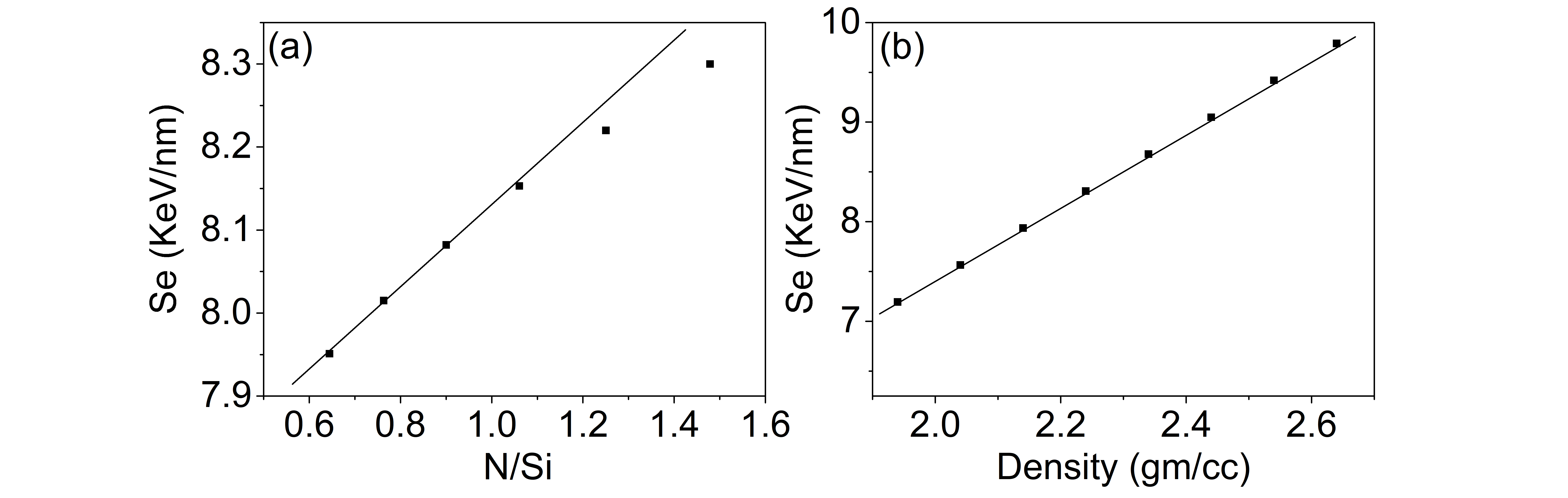}
\caption{\label{Fig2}Dependence of the electronic energy loss on the (a) Si/N ratio and (b) the film density as obtained from TRIM.}
\end{figure*}

Figure \ref{Fig1}b shows the thickness normalised hydrogen concentration for the as-deposited 
\begin{table*}[hb]
\caption{\label{Tab2}Evolution of film density and the electronic energy loss (S$ _{e} $) with irradiation fluence. The S$ _{e} $ values are determined from TRIM for the composition and density of a given thin film. The sample labelled B is the sample considered in the present work, while R and G are films reported previously\cite{bommali2015}. *Note that for R and G the contribution of the Hydrogen concentration to density has been ignored due unavailaibility of ERDA data for the irradiated films of this series (see Table I) }
\begin{tabularx}{\textwidth}{ p{2.3cm}  p{2cm}  p{2cm}  p{2cm}  p{1cm}  p{2cm}  p{2cm} p{2cm}}
\hline 
Fluence & \multicolumn{3}{c}{Density (gm/cc)} & {} & \multicolumn{3}{c}{Electronic energy loss S$ _{e} $,(KeV/nm)} \\ 
\cline{2-4} \cline{6-8} 
(ions/cm$^2$)& B &  G  & R & {} & B & G & R \\ 
\hline \hline
0 & 2.24 & 2.21 &1.9& {} & 8.2 & 7.32 & 6.51 \\ 
$5\times{10^{12}}$ & 2.19 & 2.15 & 1.87 & {}& 8 & 7.17 & 6.21 \\ 
$1\times{10^{14}}$ & 2.45 & 2.33 &2.27 & {} & 8.6 & 7.77 & 7.54 \\ 
\hline 
\end{tabularx} 
\vspace{6pt}
\end{table*}
and irradiated films. Under the aforementioned assumption the atomic concentrations of Si, N and H the films at each stage of irradiation can be combined with the corresponding thickness of the film to determine the evolution of the film density with fluence. The densities calculated in this way are presented in Table \ref{Tab2}. Also, for these specific densities and compositions, the S$ _{e} $ values are calculated using TRIM  and depicted in Table  \ref{Tab2}. Further, the general dependence of S$ _{e} $ on the composition is understood from TRIM simulation depicted in figure \ref{Fig2}a. In obtaining the above data from TRIM the target composition has been varied while keeping the density fixed at the density of the as-deposited film B.
Further, in a similar way in figure \ref{Fig2}b the variation of the S$ _{e} $ with the film density is obtained from TRIM, this time by keeping the composition fixed at the composition of as-deposited film B . Thus as figure \ref{Fig2} shows the electronic energy deposition in the target film is expected to increase monotonically with increasing N/Si ratio, and with increase in density of the film. Thus, it is quite clear that the electronic energy loss for \textit{a}-SiN$_\mathrm{\textit{x}}$:H films  is not constant but rather increases in the course of irradiation. 
\begin{figure*}[t]
\centering
\includegraphics[scale=0.7]{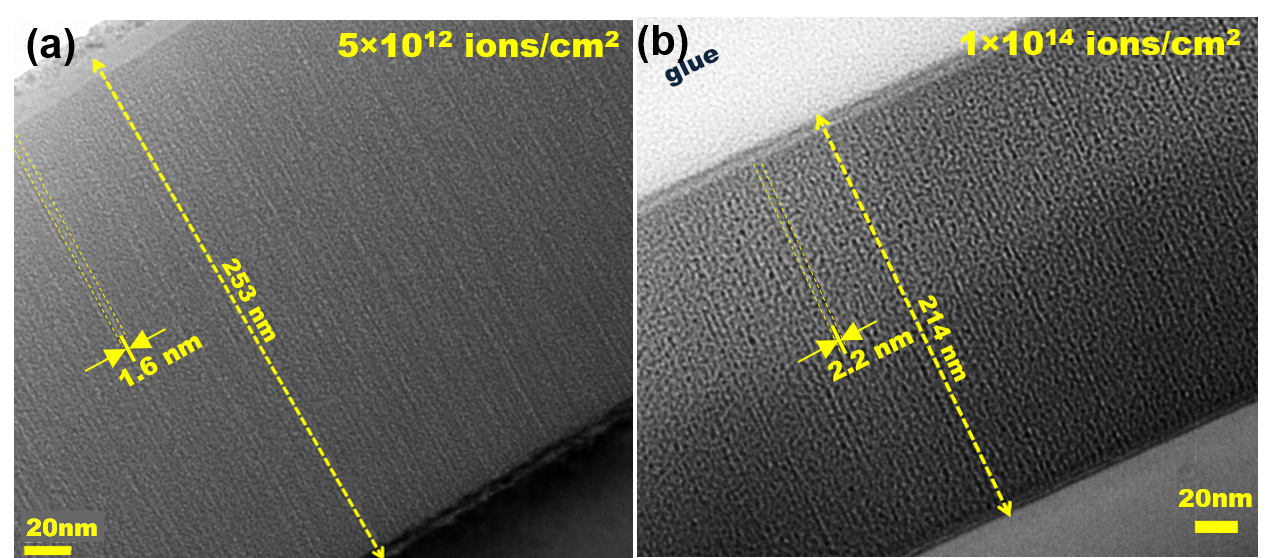}
\caption{\label{Fig3}The XTEM images of \textit{a}-SiN$_\mathrm{\textit{x}}$:H thin film irradiated with 100 MeV Ni ions showing the formation of ion tracks (a) continuous ion tracks at the fluence of $5\times{10^{12}}$ ions/cm$^2$ and (b) track discontinuities forming at the higher fluence of  $1\times{10^{14}}$ ions/cm$^2$.}
\end{figure*}
	Figure \ref{Fig3} shows the post-irradiation XTEM micrographs obtained for the thin film B. In order to interpret the results, we carry the discussion further with the understanding that the brighter regions represent sparse density, and are therefore understood to be the ion track core. The identification of the less dense regions with the core is in agreement with previous reports on \textit{a}-SiN$_\mathrm{\textit{x}}$:H and Si$ _{3}$N$_{4} $ \cite{murzalinov2018}$^{,}$\cite{van2020}$^{,}$\cite{kitayama2015}$^{,}$\cite{mota2018} wherein in the case of silicon nitride the ion tracks have an under-dense core and an over-dense shell. Similarly, the regions of dark contrast are regions of higher density and hence understood to indicate 
the track walls. The following points may be noted: Figure \ref{Fig3}a shows the formation of continuous ion tracks at the fluence of $5\times{10^{12}}$ ions/cm$^2$. The average core width of these ion tracks was measured to be ~1.6 nm across in diameter. Further irradiation to higher fluences of $5\times{10^{12}}$ ions/cm$^2$ leads to enhanced contrast between the denser and the sparser regions and visibly increased discontinuities at the ion track walls. Also visible is the widening of the track core to an average width of ~2.2 nm. The appearance of fragmented ion tracks at $1\times{10^{14}}$ ions/cm$^2$ is compatible with recent reports \cite{van2020} on polycrystalline Si$ _{3} $N$ _{4} $.
	Considering only the present thin film B (i.e. of fixed composition N/Si), we arrive at the following understanding of the microstructural evolution. As the irradiation advances the film continuously lose hydrogen, resulting in continuous densification of the film. The increasing density in turn leads to increased S$ _{e} $ (Figure \ref{Fig2}b), and the consequence of which in this case, appears as an increase in the track core diameter. The result is in agreement with reports of \textit{Vlasukova et al}. \cite{vlasukova2014} and \textit{Kitayama et al}.\cite{kitayama2015}, wherein the track core width is found to increase with the S$ _{e} $ for monoatomic ion species for irradiation. Similarly, in the past  \textit{Meftah et al.} \cite{meftah1993} have also estimated the effective track radius to increase with increasing S$ _{e} $. Further, they also studied the effect of ion velocity, wherein they found that at a given energy loss value high velocity ions lead to continuous ion tracks, lower velocity ions on the other hand lead to discontinuous ion tracks. One may therefore attribute the discontinuous ion track structures, evidenced in the present work, to decreased ion velocity in the densified medium at high fluence. Thus going by these foregoing observations, one then anticipates the tracks to widen further to dissolve/overlap, with increase in fluence. The dissolution of the ion tracks at $1\times{10^{14}}$ ions/cm$^2$ has been evidenced previously for the films\cite{bommali2015} R and G . 
\begin{figure*}[t]
\centering
\includegraphics[scale=0.45]{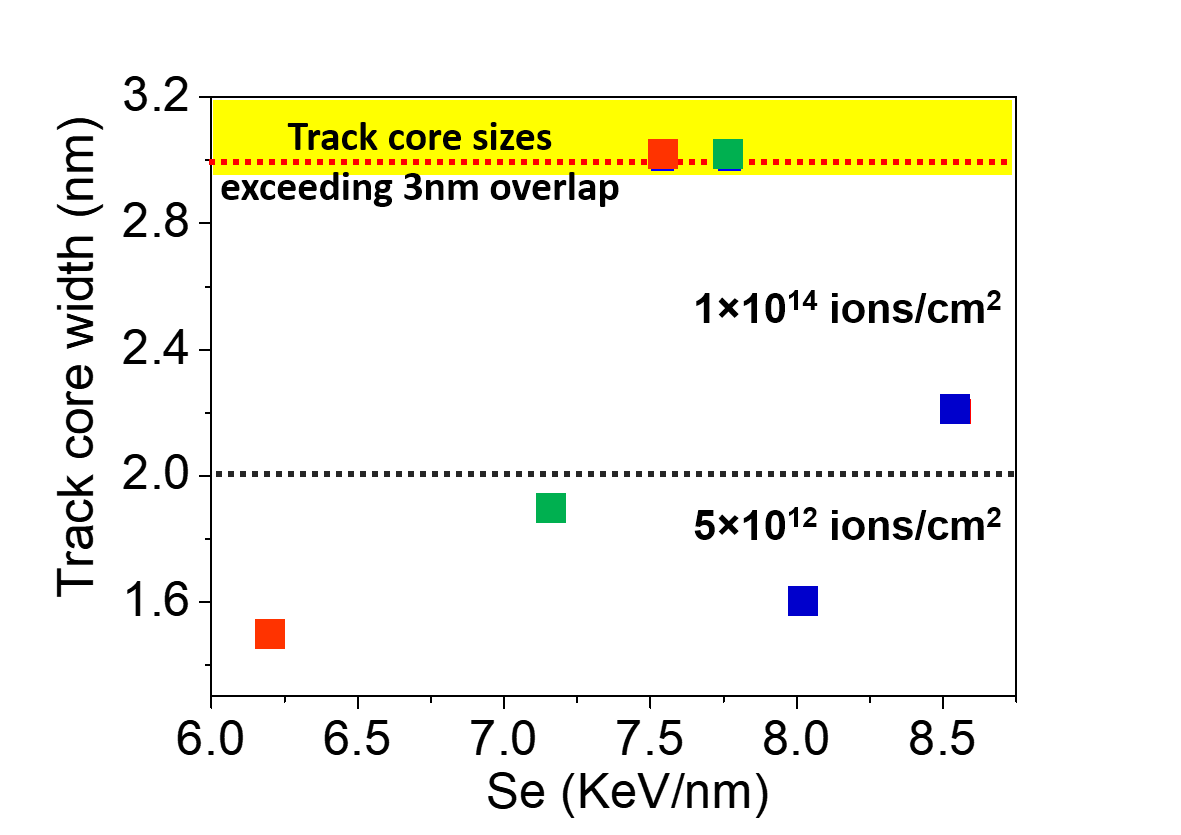}
\caption{\label{Fig4}Experimentally observed track sizes for different electronic energy loss values. Data points presented in blue correspond to present work, those in red and green correspond to R and G thin films from our previous publication. The track widths above 2 nm are observed at $1\times{10^{14}}$ ions/cm$^2$. Track widths below 2 nm are observed at $5\times{10^{12}}$ ions/cm$^2$. The highlighted region represents the track overlap regime. In order to overlap, the track-cores are assumed to be at least twice as wide as the smallest ion track thickness evidenced in this figure.}
\end{figure*}
Figure \ref{Fig4} depicts the track core widths with the S$ _{e} $ of the films (Table II) at various fluence values. The red and green data points in figure \ref{Fig4} correspond to R and G films respectively from our previous work and the data in blue corresponds to the film B in the present work. It must be noted that the track widths less than 2 nm in figure \ref{Fig4} are obtained at $5\times{10^{12}}$ ions/cm$^2$ and those above 2 nm are obtained at $1\times{10^{14}}$ ions/cm$^2$. The points at the track size of 3 nm (yellow highlight) are assumed to be the minimum tracks widths required for complete dissolution of the tracks observed in our previous work. This minimum track overlap width is taken to be as the twice of the smallest track width (1.5 nm) evidenced in previous work. This figure clearly shows that for a given film the core width increases with Se, further the core widths do not show any systematic dependence as one moves from films R to B. These differences show that the track core width is not a simple function of the energy loss (S$ _{e} $) only, and hints towards a role of other factors. The possible factors determining track morphology are understood by noting that, the ion track morphology is dependent on how the deposited energy (S$ _{e} $) is distributed radially outward \cite{kamarou2006} from the ion’s path in matter. The distribution of energy determines the energy density reached in a certain region, and consequently its morphological evolution. A higher energy density, for example, would result in a greater damage, to the extent of local melting or vaporization of the medium. Now, it is easily appreciated that the lattice/electronic thermal conductivities and electron phonon coupling of the medium, are the factors that would determine the distribution of the deposited energy. The interplay of these aforementioned factors determine how quickly the deposited energy is dissipated from the ion axis. A slower dissipation of deposited energy leads to higher energy density confined to a narrower volume and thus resulting in a wider track core \cite{van2020}. Conversely, if deposited energy is dissipated too quickly to a wider volume, it would result in less energy density and hence less damage or a narrower core. Thus the morphological evolution of films of different compositions may be determined by material constants for the given composition [Ref]. A quantitative discussion of the dependence of the ion track morphology on these material constants in R, G and B is beyond the scope of the present work. 
\begin{figure*}[ht]
\centering
\includegraphics[scale=0.5]{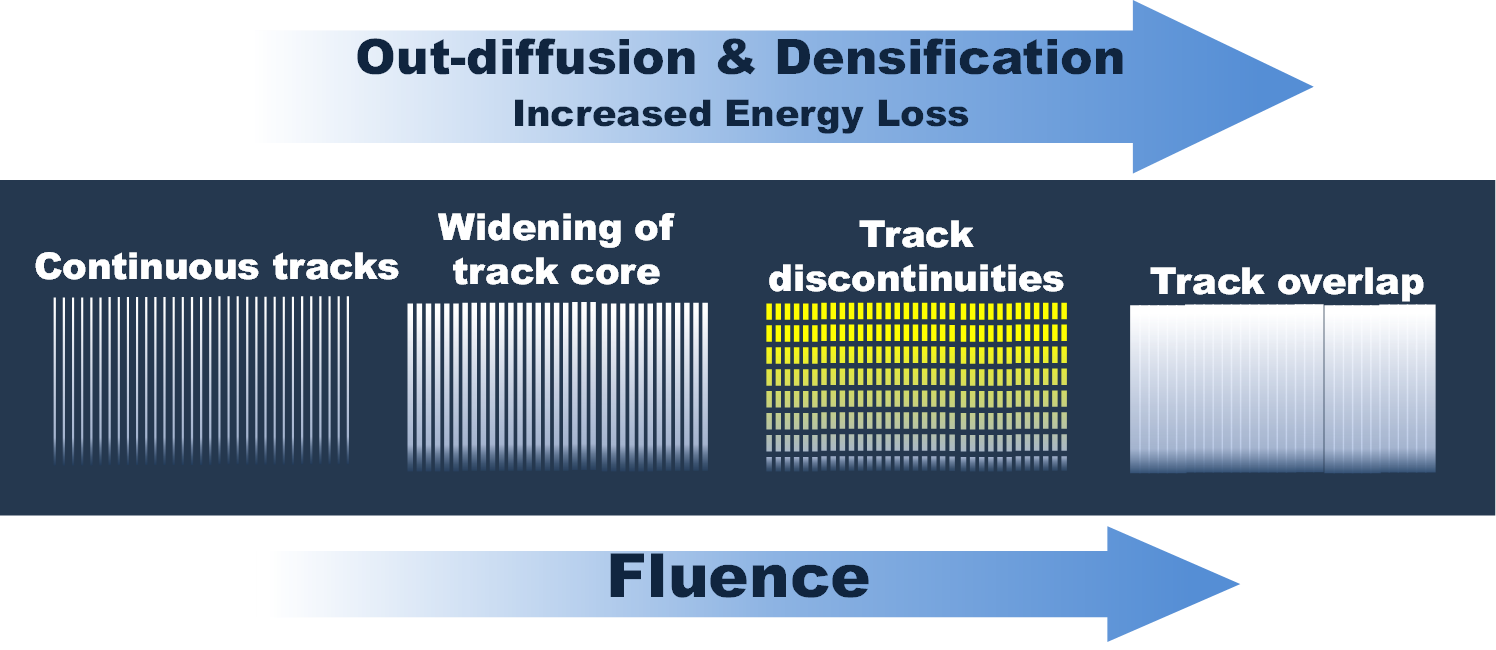}
\caption{\label{Fig5}Mechanism of ion irradiation induced morphological evolution with fluence in a-SiNx:H thin films.}
\end{figure*}
 	Thus, one arrives at the following mechanism (see figure \ref{Fig5}) for the morphological evolution within \textit{a}-SiN$_\mathrm{\textit{x}}$:H film of a given stoichiometry as irradiation progresses. Irradiation leads to out-diffusion of hydrogen and a concomitant increase in density of film, as a consequence of which the energy deposition increases due to an increase in the energy loss S$ _{e} $ is increased. The increased energy deposition leads to widening of the track cores. This processes progresses until there is coarsening of the track walls and after a threshold of electronic energy loss is reached the tracks overlap leading to  dissolution. A detailed study is under progress to see the nanotrack evolution at different initial S$ _{e} $ and various ion fluences.
\section{Conclusions} The SHI induced morphological evolution with fluence of \textit{a}-SiN$_\mathrm{\textit{x}}$:H is explained with the dynamic nature of the  electronic energy loss (S$ _{e} $). The continuous out-diffusion of hydrogen from the films, and the concomitant densification are responsible for the dynamic S$ _{e} $. The dynamic S$ _{e} $ is typical to the case \textit{a}-SiN$_\mathrm{\textit{x}}$:H and results in the formation of a discontinuous ion track morphology in the overlapping regime of $1\times{10^{14}}$ ions/cm$^2$. A composition specific model for morphological evolution is proposed. The different response of compositionally different films is attributed to the composition specific constants like thermal and lattice conductivities and electron phonon coupling constant. 
\begin{acknowledgments}
We thank the personnel at RRCAT for the XTEM measurements. Personnel at IUAC New Delhi are thanked for the ERDA measurements.
\end{acknowledgments}

\bibliography{Bommalietal_2020}

\end{document}